\documentclass{article}
\usepackage{spconf,amsmath,graphicx}

\usepackage[utf8]{inputenc}
\usepackage[T1]{fontenc}
\usepackage{hyperref}
\usepackage{url}
\usepackage{booktabs}
\usepackage{siunitx}
\usepackage{etoolbox}
\robustify\bfseries
\usepackage{amsfonts}
\usepackage{nicefrac}
\usepackage{microtype}
\usepackage{xcolor}
\usepackage{float}
\usepackage{amssymb}

\usepackage{bm}
\usepackage{tikz}
\usepackage{pgfplots}
\usepackage{subcaption}
\usepackage{glossaries}
\usepackage{placeins}
\usepackage{array}
\usepackage{multirow}

\pgfplotsset{compat=1.16}
\usetikzlibrary{pgfplots.colormaps}
\usetikzlibrary{arrows.meta}
\usetikzlibrary{3d,arrows,automata,backgrounds,calc,calendar,chains,decorations,decorations.footprints,decorations.fractals,decorations.markings,decorations.pathmorphing,decorations.pathreplacing,decorations.shapes,decorations.text,er,fadings,fit,folding,matrix,mindmap,patterns,petri,plothandlers,plotmarks,positioning,scopes,shadows,shapes.arrows,shapes.callouts,shapes,shapes.gates.logic.IEC,shapes.gates.logic.US,shapes.geometric,shapes.misc,shapes.multipart,shapes.symbols,through,topaths,trees,pgfplots.groupplots}
\usepackage{tkz-kiviat}
\usepgfplotslibrary{statistics}

\def\x{{\mathbf x}}

\def\up{\(\uparrow\)}
\def\down{\(\downarrow\) }

\title{BLINC: BLIND CALIBRATION FOR TRAINING-FREE SPEECH ENHANCEMENT ADAPTATION}
\name{Tobias Raichle, Ekaterina Gavrilko, Bin Yang}
\address{University of Stuttgart, Institute of Signal Processing and System Theory, Stuttgart, Germany}
\glsdisablehyper
\newacronym{rir}{RIR}{room impulse response}
\newacronym{se}{SE}{speech enhancement}
\newacronym{tta}{TTA}{test-time adaptation}
\newacronym{ttt}{TTT}{test-time training}
\newacronym{uda}{UDA}{unsupervised domain adaptation}
\newacronym{asr}{ASR}{automatic speech recognition}
\newacronym{snr}{SNR}{signal-to-noise ratio}
\newacronym{rtf}{RTF}{real-time factor}

\newacronym{cdf}{CDF}{cumulative distribution function}
\newacronym{pdf}{PDF}{probability density function}
\newacronym{pit}{PIT}{probability integral transform}

\newacronym{ssra}{SSRA}{self-supvervised representation based adaptation}
\newacronym{pfpl}{PFPL}{phone-fortified perceptual-loss}
\newacronym{laden}{LaDen}{latent denoising}
\newacronym{mpol}{MPol}{mask polarization}
\newacronym{diet}{DIET}{domain invariant embedding transformation}
\newacronym{tent}{tent}{test-time entropy minimization}

\newacronym{cmaes}{CMA-ES}{covariance matrix adaptation evolution strategy}

\newacronym{ss}{SS}{spectral subtraction}
\newacronym{am}{AM}{amplitude masking}

\newacronym{mse}{MSE}{mean squared error}
\newacronym{tf}{TF}{time-frequency}
\newacronym{pesq}{PESQ}{perceptual evaluation of speech quality}
\newacronym{sisdr}{SI-SDR}{scale-invariant signal-to-distortion ratio}
\glsunset{sisdr}
\newacronym{ssnr}{SSNR}{segmental signal-to-noise ratio}
\glsunset{ssnr}

\newacronym{earsw}{EARS-W}{EARS+WHAM!}
\newacronym{earsd}{EARS-D}{EARS+DEMAND}
\newacronym{vbd}{VBD}{VoiceBank+DEMAND}
\newacronym{vbw}{VBW}{VoiceBank+WHAM!}
\newacronym{dns}{DNS}{deep noise suppression}

\begin{document}
\ninept
\definecolor{mittelblau}{RGB}{0, 126, 198}
\definecolor{violettblau}{cmyk}{0.9, 0.6, 0, 0}
\definecolor{rot}{RGB}{238, 28 35}
\definecolor{apfelgruen}{RGB}{140, 198, 62}
\definecolor{gelb}{RGB}{255, 229, 0}
\definecolor{orange}{RGB}{244, 111, 33}
\definecolor{pink}{RGB}{237, 0, 140}
\definecolor{lila}{RGB}{128, 10, 145}
\definecolor{cyan}{RGB}{58, 252, 252}
\definecolor{moosgruen}{RGB}{53, 147, 134}
\definecolor{bordeaux}{RGB}{148, 23, 112}
\definecolor{kastanie}{RGB}{148, 23, 81}
\definecolor{lachs}{RGB}{255, 126, 121}
\definecolor{turkis}{HTML}{B3ECEC}
\definecolor{dunkelturkis}{HTML}{74C7EC}
\definecolor{lavendel}{RGB}{215, 131, 254}
\definecolor{hellgrau}{RGB}{224, 224, 224}
\definecolor{mittelgrau}{RGB}{128, 128, 128}
\definecolor{dunkelgrau}{RGB}{80,80,80}
\definecolor{anthrazit}{RGB}{19, 31, 31}
\definecolor{mauve}{RGB}{141, 99, 247}

\pgfplotsset{bar_1_style/.style={color=mittelblau, fill=mittelblau!50}}
\pgfplotsset{bar_2_style/.style={color=orange, fill=orange!50}}
\pgfplotsset{bar_3_style/.style={color=pink, fill=pink!50}}
\pgfplotsset{bar_4_style/.style={color=lila, fill=lila!50}}
\pgfplotsset{bar_5_style/.style={color=lavendel, fill=lavendel!50}}

\pgfplotscreateplotcyclelist{bar_default}{%
bar_1_style\\%
bar_2_style\\%
bar_3_style\\%
bar_4_style\\%
bar_5_style\\%
}

\newcommand{\PreserveBackslash}[1]{\let\temp=\\#1\let\\=\temp}
\newcommand{\newcol}{\vfill\pagebreak}

\newcolumntype{C}[1]{>{\PreserveBackslash\centering}p{#1}}

\tikzset{arrow/.style={
->,
>={Stealth[inset=0pt, angle=30:5pt]},
}}

\tikzset{arrow_large/.style={
->,
>={Stealth[inset=0pt, angle=30:12pt]},
}}

\tikzset{line_back/.style={
			color=lila,
		}}

\tikzset{arrow_back/.style={
<-,
>={Stealth[inset=0pt, angle=35:4pt]},
color=lila,
}}

\newlength{\wcol}

\newcommand{\matr}[1]{\mathbf{#1}}
\newcommand{\vect}[1]{\bm{#1}}

\newcommand{\cmark}{\ding{51}}%
\newcommand{\xmark}{\ding{55}}%
\newcommand{\glsi}[1]{\glsunset{#1}\gls{#1} (\acrlong{#1})}
\newcommand{\Glsi}[1]{\glsunset{#1}\Gls{#1} (\acrlong{#1})}

\newcommand{\dns}[1]{\(\text{DNS}_{\mathrm{#1}}\)}
\newcommand{\avg}[1]{\(\overline{\text{#1}}\)}

\newcommand{\todo}[1]{[{\color{red}TODO: }#1]}

\newlength{\corner}
\newlength{\scorner}

\setlength\belowdisplayskip{6pt}%

\tikzstyle{frozen} = [double, thick, draw=turkis, fill=turkis!50, text=turkis!330, rounded corners=\scorner]
\tikzstyle{truth} = [draw=apfelgruen, fill=apfelgruen!10, text=apfelgruen!120, rounded corners=\scorner]
\tikzstyle{trainable} = [draw=lila, fill=lila!10, text=lila!80!black, semithick, rounded corners=\scorner]
\tikzstyle{norm} = [draw=orange, fill=orange!10, text=orange!90!black, semithick]
\tikzstyle{signal} = [draw=mittelgrau, text=mittelgrau!90!black, semithick, rounded corners=\corner, inner sep=0pt]
\tikzstyle{dsp} = [draw=violettblau, fill=violettblau!20, text=violettblau!90!black, semithick, rounded corners=\scorner]

\pgfdeclarepatternformonly{south west lines}{\pgfqpoint{-0pt}{-0pt}}{\pgfqpoint{3pt}{3pt}}{\pgfqpoint{3pt}{3pt}}{
	\pgfsetlinewidth{0.4pt}
	\pgfpathmoveto{\pgfqpoint{0pt}{0pt}}
	\pgfpathlineto{\pgfqpoint{3pt}{3pt}}
	\pgfpathmoveto{\pgfqpoint{2.8pt}{-.2pt}}
	\pgfpathlineto{\pgfqpoint{3.2pt}{.2pt}}
	\pgfpathmoveto{\pgfqpoint{-.2pt}{2.8pt}}
	\pgfpathlineto{\pgfqpoint{.2pt}{3.2pt}}
	\pgfusepath{stroke}}

\pgfdeclarepatternformonly{south east lines}{\pgfqpoint{-0pt}{-0pt}}{\pgfqpoint{3pt}{3pt}}{\pgfqpoint{3pt}{3pt}}{
	\pgfsetlinewidth{0.4pt}
	\pgfpathmoveto{\pgfqpoint{0pt}{3pt}}
	\pgfpathlineto{\pgfqpoint{3pt}{0pt}}
	\pgfpathmoveto{\pgfqpoint{.2pt}{-.2pt}}
	\pgfpathlineto{\pgfqpoint{-.2pt}{.2pt}}
	\pgfpathmoveto{\pgfqpoint{3.2pt}{2.8pt}}
	\pgfpathlineto{\pgfqpoint{2.8pt}{3.2pt}}
	\pgfusepath{stroke}}

\usetikzlibrary{svg.path}

\tikzset{
	icon size/.initial = 8pt,
	icon stroke/.initial = 0.4pt,
	icon color/.initial = black,
	icon/.style = {
		draw = \pgfkeysvalueof{/tikz/icon color},
		fill = none,
		line cap = round,
		line join = round,
		line width = \pgfkeysvalueof{/tikz/icon stroke},
	},
}

\tikzset{
	pics/flame/.style = {
		code = {
			\pgfmathsetmacro\iconscale{\pgfkeysvalueof{/tikz/icon size}/24}
			\begin{scope}[xscale=\iconscale, yscale=-\iconscale, shift={(-12pt,-12pt)}]
				\draw[icon] svg {M 7.8 9.4 Q 11 7 12 3 q 2.5 5 0 10 q 3 0 5 -2.9 a 7 7 0 1 1 -9.2 -0.7 Z};
			\end{scope}
		},
	},
	pics/snowflake/.style = {
		code = {
			\pgfmathsetmacro\iconscale{\pgfkeysvalueof{/tikz/icon size}/512}
			\begin{scope}[xscale=\iconscale, yscale=-\iconscale, shift={(-256pt,-256pt)}]
				\draw[icon] svg {M 256 32 L 256 480};
				\draw[icon] svg {M 62.01 144 L 449.99 368};
				\draw[icon] svg {M 449.99 144 L 62.01 368};
				\draw[icon] svg {M 313.72 80 A 111.47 111.47 0 0 1 256 96 A 111.47 111.47 0 0 1 198.28 80};
				\draw[icon] svg {M 198.28 432 a 112.11 112.11 0 0 1 115.44 0};
				\draw[icon] svg {M 437.27 218 a 112.09 112.09 0 0 1 -57.71 -100};
				\draw[icon] svg {M 74.73 294 a 112.09 112.09 0 0 1 57.71 100};
				\draw[icon] svg {M 74.73 218 a 112.09 112.09 0 0 0 57.71 -100};
				\draw[icon] svg {M 437.27 294 a 112.09 112.09 0 0 0 -57.71 100};
			\end{scope}
		},
	},
}

\maketitle
\begin{abstract}
	\Gls{se} models degrade under domain shifts and have to adapt to unseen target domains during deployment.
	Most existing \gls{tta} methods for \gls{se} do so by adapting a subset of the model weights using a self-supervised loss, which requires backpropagation at test-time and permanently alters the model.
	We instead recalibrate the prediction itself and propose BLINC, a training-free \gls{tta} method that remaps the predicted time-frequency mask onto a bimodal target distribution by histogram matching.
	At test-time, the target distribution is parameterized from blind features of the noisy recording, so neither a reference distribution from a classical algorithm nor online metric optimization is involved.
	BLINC improves the overall quality of both evaluated \gls{se} models on almost every target condition and matches or exceeds the loss-based \gls{tta} baselines at minimal overhead.
\end{abstract}
\glsresetall
\section{INTRODUCTION}
\label{sec:intro}

Deep learning has advanced \gls{se} to striking performance, even in highly adverse noisy conditions~\cite{cmgan}.
However, this performance rests on the assumption that test conditions resemble those seen during training~\cite{laden}, which does not hold in practical deployments.
Since no training dataset can cover the full diversity of speakers and acoustic environments, \gls{se} models inevitably encounter domain shifts, causing them to degrade.
To ensure consistent performance under such shifts, models must adapt to the target domain.
For many tasks, target data is collected and annotated for this purpose.
Such annotation is impossible in \gls{se}, as the clean reference of a noisy recording cannot be restored.
Adaptation for \gls{se} must therefore be unsupervised.
\par
\Gls{tta} adapts to the target domain simultaneously with inference, using only the pre-trained model and unlabeled target data.
As opposed to \gls{uda}, \gls{tta} does not assume access to source data, which is often unavailable at deployment due to storage and privacy concerns.
Most \gls{tta} methods for \gls{se} achieve this by constructing a self-supervised loss to adapt a subset of the model's weights~\cite{laden,remixit,mpol,taap}.
However, this comes at the cost of gradient computations, needs to ramp up, permanently alters the model and can become unstable over time due to error accumulation.
\par
Instead, the prediction can be corrected in place at test-time, leaving the model weights untouched.
In-place correction requires knowing what the prediction should have been.
Existing methods achieve this by searching for the correction that maximizes a non-intrusive metric, i.e., one that takes only the estimate as input~\cite{rar}.
For mask-based \gls{se} models, we show that this can be established in advance.
\par
Under domain shifts, the predicted \gls{tf} mask loses its characteristic distribution, reducing enhancement performance~\cite{mpol}.
We find that the distribution it should follow is captured by a compact parametric family, whose parameters are predictable from blind features of the noisy recording, requiring no clean reference.
We therefore propose BLINC\footnote{Code available at \url{https://github.com/tobiaaa/SETTA}.}, which recalibrates the predicted \gls{tf} mask by histogram matching (cf. Fig.~\ref{fig:overview}).
Whereas a self-supervised loss can be optimized in a background task, direct correction must happen synchronously, so costly online metric optimization introduces significant latency.
As BLINC requires no such search at test-time, it adds minimal overhead.

\begin{figure}[h]
	\centering
	\resizebox{0.9\columnwidth}{!}{\input{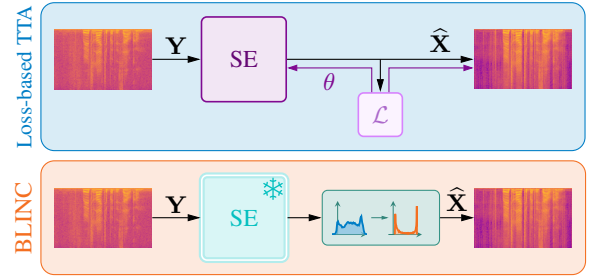}}
	\caption{Loss-based TTA minimizes a loss \(\mathcal{L}\) at test-time, adjusting either the model weights \(\theta\) or the prediction \(\widehat{\mathbf{X}}\).
		BLINC instead recalibrates the prediction of the frozen model in a single step without a loss.}
	\label{fig:overview}
\end{figure}
\vspace{-1em}
Our contributions are as follows.
\begin{itemize}
	\item We propose BLINC, a training-free \gls{tta} method that adapts mask-based \gls{se} models by histogram matching, requiring neither weight updates nor online metric optimization.
	\item We construct a signal-dependent target distribution from blind features of the noisy recording, with coefficients fitted offline against an \gls{se} metric.
	\item We analyze what the distribution calibration can and cannot repair, separating the removable calibration error from the persistent ranking error. This also explains where our approach reaches its limits.
\end{itemize}

\begin{figure*}[t]
	\centering
	\begin{minipage}[b]{0.62\textwidth}
		\centering
		\resizebox{\linewidth}{!}{\input{fig/framework.tex}}
		\caption{Overview of BLINC. The frozen SE model predicts the mask \(\matr{M}\), which \(F_\mathrm{pred}\) and \(Q_\mathrm{target}\) remap to \(\widetilde{\matr{M}}\). \(Q_\mathrm{target}\) is parameterized from blind features of \(\matr{Y}\).}
		\label{fig:framework}
	\end{minipage}
	\hfill
	\begin{minipage}[b]{0.36\textwidth}
		\centering
		\stepcounter{figure}
		\begin{subfigure}[b]{\linewidth}
			\centering
			\resizebox{!}{2.0cm}{\begin{tikzpicture}
	\begin{axis}[
			ylabel={\(Q_{\mathrm{target}}\)},
			xmin=-0.1, xmax=1.1,
			ymin=-0.15, ymax=1.15,
			height=3cm, width=5cm,
			tick label style={font=\scriptsize},
			xtick={0,0.5,1},
			xticklabels={0,\textcolor{pink}{\(b\)},1},
		]

		\addplot [
			domain=0.0:1.0,
			samples=50,
			color=mittelblau,
		]
		{max(min(1.01*(1/(1+exp(-8.5*(x-0.5)))) - 0.01, 1.0), 0.0)};

		\draw[dashed, lila] (-0.2, -0.01) -- (1.2, -0.01);

		\draw [dashed, pink] (0.5, -0.15) -- (0.5, 1.15);

		\draw[arrow, mauve] (0.65, 0.779) -- (0.75, 0.779);
		\draw[arrow, mauve] (0.65, 0.779) -- (0.55, 0.779);
		\draw[arrow, mauve] (0.35, 0.211) -- (0.25, 0.211);
		\draw[arrow, mauve] (0.35, 0.211) -- (0.45, 0.211);

		\node at (0.8, 0.779) [mauve, font=\scriptsize] {\(a\)};

		\node at (0.9, 0.10) [lila, font=\scriptsize] {\(c\)};

	\end{axis}
\end{tikzpicture}}
			\vspace{-0.5em}
			\caption{Quantile function \(Q_\mathrm{target}\)}
			\label{fig:quant}
		\end{subfigure}
		\par\vspace{0.5em}
		\begin{subfigure}[b]{\linewidth}
			\centering
			\resizebox{!}{2.0cm}{\begin{tikzpicture}
	\begin{axis}[
			ylabel={\(f_{\mathrm{target}}\)},
			xmin=-0.1, xmax=1.1,
			ymin=-0.1, ymax=5.1,
			height=3cm, width=5cm,
			tick label style={font=\scriptsize},
			yticklabel={\hphantom{.5}\pgfmathprintnumber{\tick}},
		]

		\addplot [
			domain=-0.2:1.2,
			samples=500,
			color=mittelblau,
		]
		{max(1/8.5 * 1.01/((x-1)*(-0.0-x)),0)};

	\end{axis}
\end{tikzpicture}}
			\vspace{-0.5em}
			\caption{Induced density \(f_\mathrm{target}\)}
			\label{fig:pdf}
			\vspace{-0.5em}
		\end{subfigure}
		\vspace{-1em}
		\addtocounter{figure}{-1}
		\caption{\vphantom{\(F_\mathrm{pred}\)}Parametric target \(Q_\mathrm{target}\) with floor \(c\), midpoint \(b\) and sharpness \(a\), and the density it induces.\vphantom{$\widetilde{\matr{M}}$}}
		\label{fig:mask_dist}
	\end{minipage}
	\vspace{-1.0em}
\end{figure*}

\section{RELATED WORK}
\label{sec:rel}
Most \gls{tta} methods for \gls{se} adapt the model by minimizing a self-supervised loss, and differ mainly in how that loss is constructed.
The first \gls{tta}-compatible approach to adapt \gls{se} models was RemixIT~\cite{remixit}, which uses a teacher model to construct a weakly labeled dataset that is used to train a student model.
LaDen~\cite{laden} constructs pseudo-labels and computes a loss in the embedding space of a large speech encoder.
In~\cite{taap}, the authors propose using a pre-trained clean speech prior as the target for adaptation.
In MPol~\cite{mpol}, mask-based \gls{se} models are adapted by comparing the predicted mask distribution to that of a more robust, yet low-fidelity, classical reference.
The predicted distribution is therefore pulled toward that of a suboptimal reference, bounding the adaptation.
Adapting the model in this way requires backpropagation at test-time and can become unstable as errors accumulate over long deployments.
\par
Instead, the rethink-and-refine correction module~\cite{rar}, denoted RaR, refines the prediction directly.
The predicted waveform is locally interpolated with the noisy signal to reduce over-suppression by optimizing the interpolation with respect to a non-intrusive \gls{se} metric.
The interpolation weights are chosen per speech unit, which requires an \gls{asr} model to segment the recording.
Both the segmentation and the optimization run at test-time, which introduces significant overhead and is confined to metrics that are non-intrusive and differentiable.
Moreover, interpolating toward the noisy recording can only restore over-suppressed content, and reintroduces noise proportionally to restoration.

\section{METHODOLOGY}
\label{sec:method}

We consider mask-based \gls{se}, where a noisy recording in the \gls{tf} domain is modeled as \(\matr{Y}=\matr{X}+\matr{N}\in\mathbb{C}^{F\times T}\) with clean speech \(\matr{X}\) and additive noise \(\matr{N}\).
A model predicts a real-valued mask \(\matr{M}\in\mathbb{R}^{F\times T}\) over \(F\) frequency bins and \(T\) frames that weights each bin of \(\matr{Y}\), forming the estimate \(\widehat{\matr{X}}=\matr{M}\odot\matr{Y}\), where \(\odot\) is the elementwise product.
\par
In the source domain, the predicted mask \(\matr{M}\) is bimodal, with most entries close to either full attenuation or pass-through.
Under domain shifts this bimodality erodes and mask values accumulate at intermediate levels, which degrades \gls{se} performance~\cite{mpol}.
Since the mask for each sample is available at inference, we recalibrate its distribution directly, leaving the model weights untouched.
The recalibration requires two components, a mechanism that imposes a given target distribution on the prediction (Sec.~\ref{sec:histmatch}) and the target itself (Secs.~\ref{sec:target} and~\ref{sec:distparam}).
\subsection{Histogram matching}
\label{sec:histmatch}
Imposing a target distribution on the predicted mask is a histogram matching problem, which the \gls{pit} solves in closed form~\cite{pit}.
Let \(m\) denote a generic entry of the predicted mask \(\matr{M}\).
As illustrated in Fig.~\ref{fig:framework}, applying the marginal \gls{cdf} \(F_\mathrm{pred}\) of \(\matr{M}\) to its own entries maps them to a standard uniform distribution, and a quantile function \(Q_\mathrm{target}=F^{-1}_\mathrm{target}\) maps them from there to a target distribution
\begin{align}
	\tilde{m} = Q_\mathrm{target}\bigl(F_\mathrm{pred}(m)\bigr).
	\label{eq:pit}
\end{align}
The recalibrated mask \(\widetilde{\matr{M}}\in\mathbb{R}^{F\times T}\) follows \(F_\mathrm{target}\) instead of \(F_\mathrm{pred}\).
\par
Since the total remapping \(Q_\mathrm{target} \circ F_{\mathrm{pred}}\) is non-decreasing, the \(k\)-th smallest entry remains the \(k\)-th smallest after remapping.
The recalibration therefore preserves the ranking of the entries of \(\matr{M}\) and replaces only their values, i.e., their distribution.
The ranking determines which \gls{tf} bins are attenuated more strongly than others and carries most of the structure of the prediction.
The distribution determines how much attenuation each of the bins receives and thus the calibration of the mask.
\par
As \(F_\mathrm{pred}\) is unknown, we replace it by the empirical \gls{cdf} over the \(N=F\cdot T\) entries of \(\matr{M}\).
The empirical \gls{pit} assigns the \(k\)-th smallest entry the \(k\)-th target quantile
\begin{align}
	\tilde{m}_{(k)} = Q_\mathrm{target}(k/N) \quad \forall k,
	\label{eq:emp_pit}
\end{align}
where \(\tilde{m}_{(k)}\) denotes the \(k\)-th smallest entry of \(\widetilde{\matr{M}}\).

\subsection{Bimodal target}
\label{sec:target}

Speech is spectrally more concentrated than ambient noise~\cite{speech_enhancement}, so most \gls{tf} bins at \gls{snr} \(\gtrsim 0\,\text{dB}\) are dominated by a single source.
A well-calibrated mask therefore decisively separates attenuation from pass-through, concentrating its value distribution into two modes~\cite{mpol,ibm}.
Since the remapping consumes only the quantile function of the target (Eq.~\ref{eq:pit}), we specify this target directly via \(Q_{\mathrm{target}}\) rather than through a density.
A bimodal distribution corresponds to a quantile function that is flat near its ends and steep in between, which we model with the sigmoid \(\sigma(x)=1/(1+\mathrm{e}^{-x})\),
\begin{align}
	Q_\mathrm{target}(x)=(1-c)\,\sigma\!\big(a(x-b)\big)+c.
	\label{eq:qtarget}
\end{align}
The floor is set by \(c\), the midpoint \(b\) corresponds to the fraction of bins assigned to the noise mode, and the sharpness \(a\) controls the slope, as Fig.~\ref{fig:mask_dist} illustrates.
The floor attenuates the noise mode rather than removing it, which retains speech misranked into it and avoids the artifacts of complete suppression~\cite{speech_enhancement}.
\par
The bimodal shape fixes the family of the target \(Q_\mathrm{target}\) but leaves its parameters open.
A single quantile function cannot serve every recording, since the exact distribution of a well-calibrated mask differs between signals.
The midpoint, for instance, varies with the amount of speech the recording contains.
We therefore construct a target distribution for each input signal.

\subsection{Blind parameter estimation}
\label{sec:distparam}

We propose that the calibration target can be constructed from simple, blind features extracted from the respective noisy signal, that is, features that require no clean reference.
Since the midpoint \(b\) is the fraction of bins in the noise mode, its natural feature is the speech-activity fraction \(P_\mathrm{Sp}\)~\cite{blind_feat}.
The floor and the sharpness balance residual noise against lost speech, and that balance also shifts with the speech content, so we model all three parameters, \(a, b\) and \(c\), as affine functions of \(P_\mathrm{Sp}\).
\par
\begin{figure}[ht]
	\centering
	\resizebox{0.7\columnwidth}{!}{\begin{tikzpicture}
	\begin{axis}[
			xlabel={\(P_{\mathrm{Sp}}\)},
			ylabel={\(b\)},
			height=4cm, width=6cm,
			clip mode=individual,
            xmin=0.14, xmax=0.65,
            ymin=0.05,
		]

		\addplot [
			only marks,
			mark size=0.5,
			mark=*,
			mittelblau,
		]
		table [x=x, y=y]
			{fig/data/s_vs_act_frac.dat};
	\end{axis}
\end{tikzpicture}}
	\vspace{-0.75em}
	\caption{Signal-wise optimal midpoint \(b\) over the blind speech-activity fraction \(P_\mathrm{Sp}\).}
	\label{fig:param_reg}
	\vspace{-0.5em}
\end{figure}
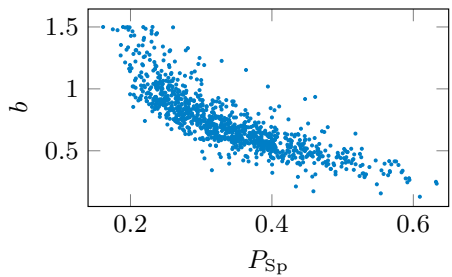

Deriving the coefficients of these affine functions requires an objective for the calibration.
For a correct ranking, an optimal calibration could attenuate exactly the noise in each bin, restoring the clean magnitude.
In practice the ranking contains errors, so every calibration passes noise or removes speech, and different use cases weigh these errors differently.
Optimality is therefore relative to a criterion, for which we use an \gls{se} metric.
Following~\cite{cmgan}, we use PESQ~\cite{pesq}, so the optimal target distribution for a signal is the one that yields the highest PESQ for that signal.
Fig.~\ref{fig:param_reg} illustrates the relationship between \(P_\mathrm{Sp}\) and the optimal midpoint \(b\).
\par
We fit the coefficients offline by maximizing the criterion averaged over a validation dataset.
As there are only six coefficients, a black-box search suffices, so the criterion need not be differentiable.
\par
At test-time, \(P_\mathrm{Sp}\) is computed from the noisy mixture and mapped to \((\hat a,\hat b,\hat c)\), which define \(Q_\mathrm{target}\) through Eq.~\ref{eq:qtarget} and thus the remapping in Eq.~\ref{eq:emp_pit}.
The target \(Q_\mathrm{target}\) contains no scale parameter, since a global scale is unidentifiable under scale-invariant criteria like PESQ, although it does affect signal-level metrics.
We therefore rescale the enhanced waveform to the estimated speech power
\begin{align}
	P_\mathrm{out} = \frac{\mathrm{SNR}}{1+\mathrm{SNR}}\,P_\mathrm{in},
\end{align}
where \(P_\mathrm{in}\) is the power of the noisy mixture and the \(\mathrm{SNR}\) is estimated blindly according to~\cite{blind_feat} for each recording.

\section{EXPERIMENTS}
\label{sec:exp}
\subsection{Datasets}
\label{sec:data}

Following~\cite{laden,mpol}, we train the source \gls{se} models on \gls{earsw}~\cite{ears,wham} and evaluate the \gls{tta} methods on various target domains to isolate the effect of different domain shifts.
\Gls{earsd}~\cite{ears,demand} retains the source speech corpus and replaces the noise, whereas \gls{vbw}~\cite{voicebank,wham} retains the noise and replaces the speech.
\Gls{vbd}~\cite{voicebank,demand,vbd} replaces both, and the \gls{dns} challenge test set~\cite{dns_2022} additionally covers six languages.
\par
Since the optimal calibration depends on how reliable the underlying ranking is (cf. Sec.~\ref{sec:distparam}), coefficients fitted on the source domain would be tuned to an unrealistically accurate ranking.
We therefore fit the coefficients on a separate validation domain, for which we mix the studio recordings of DAPS~\cite{daps} with the ambient noise of TAU~\cite{tau}.
Neither corpus overlaps with any of the target domains and the coefficients are applied unchanged throughout.

\subsection{Models}
\label{sec:models}

Since BLINC recalibrates the predicted mask, it applies to mask-based \gls{se} models, which cover a large share of current architectures.
To test whether the method transfers across architectures, we evaluate on two different models.
To represent simple architectures, we use a lightweight \gls{am} model~\cite{laden} trained with an MSE loss.
CMGAN~\cite{cmgan} represents the current state of the art with an explicit perceptual focus and refines its magnitude mask with a complex additive term, so it is mask-based only in a wider sense.
\par
The six coefficients of Sec.~\ref{sec:distparam} are fitted separately for each model, as the models differ in ranking reliability under domain shifts and the additive term changes the role of the CMGAN mask.
\subsection{Experimental details}

Perceptual quality is measured by PESQ~\cite{pesq} and the composite measures CSIG, CBAK and COVL~\cite{mos}, signal-level quality by \gls{ssnr} and \gls{sisdr}.
\par
We compare against the unadapted source model and against RemixIT~\cite{remixit}, LaDen~\cite{laden} and MPol~\cite{mpol}, using the hyperparameters reported by their authors.
All \gls{tta} methods adapt the same source checkpoints.
No complete official implementation of RaR~\cite{rar} is available, so we re-implement it following the description in the paper.
As the re-implementation cannot be validated against the original, RaR is compared on computational cost only, which depends on the steps of the method rather than on their exact tuning.
Tab.~\ref{tab:rtf} shows the \glspl{rtf} of the compared methods evaluated on an Nvidia RTX Pro 6000 GPU.
As BLINC does not require backpropagation, parameter updates or online metric optimization, its computational overhead is negligible.
\begin{table}[ht]
	\centering
	\vspace{-0.5em}
	\caption{\Gls{rtf}\down with the AM model on VBD. Factors \(\leq1\) indicate real-time capability.}
	\vspace{-0.5em}
	\label{tab:rtf}
	\begin{tabular}{c|cccccc}
	\toprule
	Source & RemixIT & LaDen & MPol  & RaR   & BLINC          \\
	\midrule
	0.003  & 0.014   & 0.372 & 0.032 & 0.903 & \textbf{0.003} \\
	\bottomrule
\end{tabular}

	\vspace{-1.5em}
\end{table}

\begin{table*}[ht]
	\centering
	\caption{Evaluation results averaged over the target datasets. SSNR and SI-SDR in dB.}
	\vspace{-0.5em}
	\label{tab:results}
	\begin{tabular}{ll*{5}{S[table-format=1.2]}S[table-format=2.2]}
	\toprule
	                                       &         & {\avg{PESQ} \(\uparrow\)} & {\avg{CSIG} \(\uparrow\)} & {\avg{CBAK} \(\uparrow\)} & {\avg{COVL} \(\uparrow\)} & {\avg{SSNR} \(\uparrow\)} & {\avg{SI-SDR} \(\uparrow\)} \\
	\midrule
	\multirow{5}{*}{\rotatebox{90}{AM}}    & Source  & 2.05                      & 3.07                      & 2.78                      & 2.52                      & 7.42                      & 12.28                       \\
	                                       & RemixIT & 2.06                      & 3.10                      & 2.80                      & 2.54                      & \bfseries 7.48            & \bfseries 12.47             \\
	                                       & LaDen   & 2.13                      & 3.13                      & 2.80                      & 2.59                      & 7.01                      & 12.33                       \\
	                                       & MPol    & 2.10                      & 3.17                      & \bfseries 2.82            & 2.60                      & 7.40                      & 12.35                       \\
	                                       & BLINC   & \bfseries 2.17            & \bfseries 3.21            & 2.78                      & \bfseries 2.66            & 6.19                      & 11.50                       \\
	\midrule
	\multirow{5}{*}{\rotatebox{90}{CMGAN}} & Source  & 2.60                      & 3.75                      & 3.02                      & 3.15                      & 6.00                      & 11.32                       \\
	                                       & RemixIT & 2.60                      & 3.77                      & 3.03                      & 3.17                      & 5.92                      & 11.52                       \\
	                                       & LaDen   & 2.62                      & \bfseries 3.81            & \bfseries 3.07            & 3.20                      & \bfseries 6.31            & \bfseries 12.09             \\
	                                       & MPol    & 2.64                      & 3.78                      & \bfseries 3.07            & 3.20                      & 6.27                      & 11.78                       \\
	                                       & BLINC   & \bfseries 2.66            & \bfseries 3.81            & 3.02                      & \bfseries 3.22            & 5.17                      & 9.42                        \\
	\bottomrule
\end{tabular}

	\vspace{-1.0em}
\end{table*}
\section{RESULTS}
\label{sec:results}

\begin{figure}[ht]
	\centering
	\begin{subfigure}[b]{\columnwidth}
		\centering
		\resizebox{0.7\columnwidth}{!}{\begin{tikzpicture}
	\tkzKiviatDiagram[scale=0.7,label distance=0.0cm,
		gap=1.0,
		ymin=-0.1,
		ymax=0.2,
		label space=3.75cm,
		lattice=3]{{~~~~~~EARS-D},VBD,VBW,\dns{EN},{\dns{GE}\hspace{2em}~},{\dns{IT}\hspace{2em}~},\dns{RU},\dns{SP},\dns{FR}}
	\tkzKiviatLine[thick,color=mittelblau,mark=none,opacity=.5](
	0.0,
	0.0,
	0.0,
	0.0,
	0.0,
	0.0,
	0.0,
	0.0,
	0.0
	)
	\tkzKiviatLine[thick,color=orange,opacity=.5](
	-0.0332,
	-0.0039,
    -0.0013,
	0.0343,
	0.0325,
	0.0367,
	-0.0156,
	0.0319,
	0.0340
	)
	\tkzKiviatLine[thick,mark size=4pt,color=lila](
	-0.0751,
	0.131,
	0.0488,
	0.0696,
	0.0952,
	0.0552,
	0.0575,
	0.0984,
	0.1083
	)
	\tkzKiviatLine[thick,mark size=4pt,color=pink](
	0.0334,
	0.0789,
    0.1005,
	0.0691,
	0.0995,
	0.0802,
	0.0628,
	0.0941,
	0.0924
	)

	\tkzKiviatLine[thick,mark size=4pt,color=lavendel](
	0.0280,
	0.1366,
	0.1328,
	0.1292,
	0.1666,
	0.1500,
	0.1274,
	0.2000,
	0.2011
	)

	\tkzKiviatGrad[prefix=,unity=1,suffix=](0)
\end{tikzpicture}}
		\caption{AM}
		\label{fig:res_am}
	\end{subfigure}
	\begin{subfigure}[b]{\columnwidth}
		\centering
		\resizebox{0.7\columnwidth}{!}{\begin{tikzpicture}
	\tkzKiviatDiagram[scale=0.7,label distance=0.0cm,
		gap=1.0,
		ymin=-0.1,
		ymax=0.2,
		label space=3.75cm,
		lattice=3]{{~~~~~~EARS-D},VBD,VBW,\dns{EN},{\dns{GE}\hspace{2em}~},{\dns{IT}\hspace{2em}~},\dns{RU},\dns{SP},\dns{FR}}
	\tkzKiviatLine[thick,color=mittelblau,mark=none,opacity=.5](
	0.0,
	0.0,
	0.0,
	0.0,
	0.0,
	0.0,
	0.0,
	0.0,
	0.0
	)
	\tkzKiviatLine[thick,color=orange,opacity=.5](
	0.0048,
	-0.0094,
	0.0582,
	0.0530,
	0.1423,
	0.0927,
	0.0290,
	0.0646,
	0.1192
	)
	\tkzKiviatLine[thick,mark size=4pt,color =lila](
	0.0243,
	-0.0441,
	0.0495,
	0.0794,
	0.1709,
	0.1800,
	0.0418,
	0.1339,
	0.1627
	)

	\tkzKiviatLine[thick,mark size=4pt,color=pink](
	0.0071,
	-0.0086,
	0.0254,
	0.0724,
	0.1788,
	0.1658,
	0.0447,
	0.1184,
	0.1601
	)

	\tkzKiviatLine[thick,mark size=4pt,color=lavendel](
	0.1826,
	0.1075,
	0.1976,
	0.0122,
	0.1423,
	0.1292,
	-0.0370,
	0.1435,
	0.1048
	)

	\tkzKiviatGrad[prefix=,unity=1,suffix=](0)
\end{tikzpicture}}
		\caption{CMGAN}
		\label{fig:res_cmgan}
	\end{subfigure}
	\par\vspace{0.3em}
	\resizebox{0.9\columnwidth}{!}{\begin{tikzpicture}
	\begin{axis}[%
			hide axis,
			xmin=10,
			xmax=50,
			ymin=0,
			ymax=0.4,
			legend style={font=\small, legend columns=-1},
			legend cell align={left},
		]
		\addlegendimage{mittelblau, thick}
		\addlegendentry{Source \ }
		\addlegendimage{orange, thick}
		\addlegendentry{RemixIT \ }
		\addlegendimage{lila, thick}
		\addlegendentry{LaDen}
		\addlegendimage{pink, thick}
		\addlegendentry{MPol}
		\addlegendimage{lavendel, thick}
		\addlegendentry{BLINC}
	\end{axis}
\end{tikzpicture}}
	\caption{\(\Delta\)COVL per target domain relative to the source model.}
	\label{fig:res}
	\vspace{-1.5em}
\end{figure}
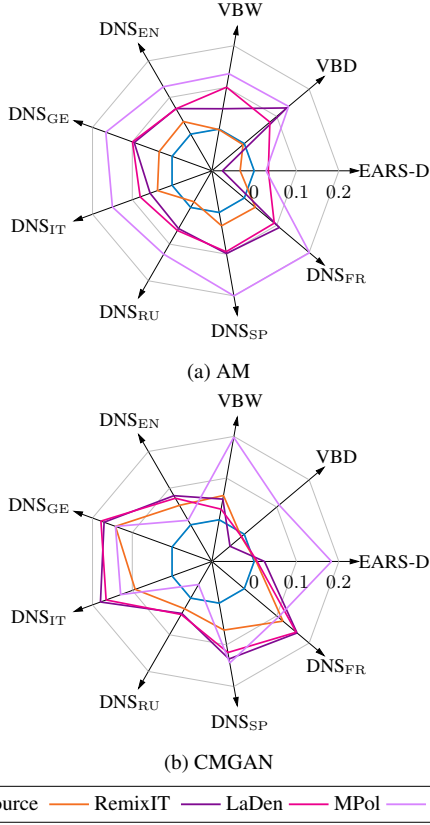
BLINC achieves the highest PESQ, CSIG and COVL of all methods on both models in Tab.~\ref{tab:results}, at nearly the \gls{rtf} of the source model.
As PESQ was used as the criterion to derive the coefficients of Sec.~\ref{sec:distparam}, the calibration retains speech at the cost of residual noise, which raises CSIG, leaves CBAK flat and sacrifices signal-level quality.
\par
On the \gls{am} model, Fig.~\ref{fig:res} shows a gain on every target condition, which is the largest of all methods on all but \gls{earsd} and grows with the severity of the shift.
On CMGAN, BLINC leads on \gls{earsd}, \gls{vbw} and \gls{vbd}, where the loss-based baselines gain little or lose, but trails LaDen and MPol on most of the \gls{dns} conditions and falls below the source model on one of them.
Since the remapping removes calibration error only, this suggests that the \gls{dns} shift also degrades the ranking of the stronger model, which recalibration cannot repair.
As CMGAN complements its mask with an additive term, the gains from recalibrating the mask alone suggest that miscalibration is a general failure mode of \gls{tf} masking under domain shift.

\subsection{Ablation study}
\label{sec:ablation}

\begin{table}[ht]
	\centering
	\caption{Ablation study with the AM model on VBD. The lower block uses information unavailable at test-time. SI-SDR in dB.}
	\vspace{-0.5em}
	\label{tab:ablation}
	\begin{tabular}{lcc}
	\toprule
	Target           & PESQ\up       & SI-SDR\up      \\
	\midrule
	Source           & 2.42          & 11.49          \\
	Constant target  & 2.45          & 10.73          \\
	BLINC            & \textbf{2.55} & 11.24          \\
	BLINC, SI-SDR    & 2.29          & \textbf{12.21} \\
	\midrule
	BLINC, in-domain & 2.65          & \textbf{11.02} \\
	Sigmoid oracle   & 2.75          & \textbf{11.02} \\
	Free-form oracle & \textbf{2.79} & 10.93          \\
	\bottomrule
\end{tabular}

	\vspace{-1.0em}
\end{table}

Tab.~\ref{tab:ablation} relaxes the construction of the target, from a signal-independent one to oracles that use ground-truth information.
A constant target retains the bimodal shape but applies the same quantile function to every recording.
It achieves a small perceptual gain over the source model at a large signal-level cost.
The signal-dependent target of BLINC improves on it in both metrics, gaining more perceptual quality while recovering most of the signal-level loss.
\par
Fitting the coefficients to \gls{sisdr} instead of PESQ reverses the trade, improving signal-level performance over the source model at a perceptual cost.
This illustrates the impact of the calibration criterion.
\par
As described in Sec.~\ref{sec:data}, the optimal calibration parameters depend on the model degradation and thus on the domain.
Fitting the coefficients on the target domain achieves further perceptual gain, quantifying the cost of transferring them from the validation domain.
\par
The last two rows of Tab.~\ref{tab:ablation} compute the optimal quantile function for each recording, either from the sigmoid family of Sec.~\ref{sec:target} or as a free-form piecewise linear function.
Against the sigmoid oracle, the in-domain fit shows what is lost by modeling the parameters as affine functions of \(P_\mathrm{Sp}\).
The sigmoid oracle comes close to the free-form oracle, indicating that the sigmoid family is sufficiently expressive.
The free-form oracle approximates the ceiling of any rank-preserving correction, and the remaining error is attributable to ranking error.

\section{CONCLUSION}
\label{sec:conc}
We presented BLINC, a training-free \gls{tta} method for \gls{se} that recalibrates the predicted mask in place and matches or exceeds the loss-based baselines at nearly the cost of the source model alone.
Our key finding is that the distribution the mask should follow is known in advance, since a compact parametric family captures it and blind features of the noisy recording predict its parameters.
Histogram matching onto this target leaves the model untouched and applies to any mask-based \gls{se} model.
The gains on CMGAN suggest that miscalibration is a general failure mode of \gls{tf} masking under domain shift.
The distinction between calibration and ranking error explains where these gains saturate and where weight adaptation retains an advantage.
By removing the online metric search, this work makes in-place correction a practical alternative to weight adaptation where backpropagation is prohibitive.
\par
The focus of the calibration follows the criterion, and fitting against PESQ incentivizes perceptual quality at a signal-level cost.
Fitting against several criteria jointly or regularizing the coefficients could enable a broader focus.
Future work should explore the use of additional blind features to improve the blind parameter estimation.
Additionally, extending to corruptions beyond additive noise, such as reverberation, is left for future research.

\FloatBarrier

\bibliographystyle{IEEEbib}
\bibliography{refs}

\end{document}